\documentclass{Interspeech2024}

\interspeechcameraready

\title{Exploring compressibility of transformer based text-to-music (TTM) models}

\name[affiliation={2}]{Vasileios}{Moschopoulos}
\name[affiliation={2}]{Thanasis}{Kotsiopoulos}
\name[affiliation={1}]{Pablo}{Peso Parada}
\name[affiliation={2}]{Konstantinos}{Nikiforidis}
\name[affiliation={3}]{Alexandros}{Stergiadis}
\name[affiliation={2}]{Gerasimos}{Papakostas}
\name[affiliation={1}]{Md Asif}{Jalal}
\name[affiliation={1}]{Jisi}{Zhang}
\name[affiliation={2}]{Anastasios}{Drosou}
\name[affiliation={1}]{Karthikeyan}{Saravanan}

\usepackage{algorithm} %
\usepackage{algpseudocode} %
\usepackage{cite}
\algtext*{EndFor}%
\algtext*{EndIf}%

\address{
  \small
  $^1$Samsung R\&D Institute UK (SRUK), United Kingdom\\
  $^2$Centre for Research and Technology Hellas, Greece\\
  $^3$Pragma IoT SA, Hellas, Greece
}
\email{} %


\renewcommand{\footnotesize}{\scriptsize} %
\begin{document}

\maketitle

\begin{abstract}
State-of-the art Text-To-Music (TTM) generative AI models are large and require desktop or server class compute, making them infeasible for deployment on mobile phones.
This paper presents an analysis of trade-offs between model compression and generation performance of TTM models. We study compression through knowledge distillation and specific modifications that enable applicability over the various components of the TTM model (encoder, generative model and the decoder).
Leveraging these methods we create TinyTTM (89.2M params) that achieves a FAD of 3.66 and KL of 1.32 on MusicBench dataset, better than MusicGen-Small (557.6M params) but not lower than MusicGen-small fine-tuned on MusicBench.

\end{abstract}

\section{Introduction}
Recent developments in generative AI have facilitated the emergence of various novel applications, including text-to-music (TTM) systems to generate music content from textual prompts. These systems allow users to efficiently craft music compositions that fulfill their specific requirements, reducing the necessity for expertise in music composition. Recent works on this topic, such as MusicGen \cite{musicGen} or MusicLDM \cite{musicldm}, have demonstrated the capability of generating high-quality music samples, highlighting the potential of TTM.

TTM models consist of three primary components: a text encoder, a generative model and a decoder. The text encoder is responsible for translating the input text into an embedding representation.
Different text encoders are employed in the literature such as a contrastive language-audio pretrained model based on Mulan \cite{mulan} or CLAP \cite{laionclap} employed in  MusicLM \cite{musicLM}, Stable Audio\cite{stableAudio}, AudioLDM\cite{audioldm} and AudioLDM2 \cite{audioldm2} or language model based \cite{raffel2020exploring,flanT5} integrated in MAGNeT \cite{magnet}, MusigGen \cite{musicGen}, Mustango \cite{mustango} and AudioLDM2 \cite{audioldm2}.
There are two primary approaches used for the generative model: autoregressive/nonautoregressive transformers as in \cite{musicLM,musicGen,magnet} where the transformer produces a sequence of discrete tokens conditioned to the encoder embedding via the cross-attention mechanism; and diffusion based models \cite{audioldm,audioldm2,musicldm,stableAudio,mustango} which learn to denoise a randomly generated latent space conditioned to the encoder embedding. 
The decoder, accepts the generative model output and transforms it into a time-domain signal via audio codecs \cite{musicLM,musicGen,magnet,audioldm, audioldm2, musicldm, mustango,stableAudio}.

Contemporary TTM models \cite{musicGen, audioldm} are large with 100s of millions of parameters.
Even the smaller variants of \cite{musicGen}, and \cite{audioldm} have 557.6M (authors mention 300M parameters but this refers to autoregressive transformer only), and 428M parameters respectively. It is essential to minimise the size of such models to deploy on resource-constrained devices.
In this work we propose and analyse different methods to reduce the size of each component in TTM. 
A popular approach towards reducing model size is knowledge distillation, where the ``knowledge'' of the original model often referred to as the ``teacher'' is distilled into a smaller target size model referred to as the ``student''. 
This is typically quantified with the Kullback-Leibler divergence loss \cite{ase_detection,KD_multilevel_emo}, which measures the discrepancy between teacher and student output distributions. Furthermore, the intermediate layer outputs of the teacher model can be also exploited \cite{kd_tinybert,KD_shao2023whisperkdq,KD_PatientKD,su_ernie-tiny_2021}, providing the student model with access to deeper knowledge and encouraging finer alignment between the models' internal representations. Another common strategy is weight transfer \cite{kd_literature_review_main}, where part of the teacher weight matrices are copied directly to the student model \cite{kd_distilbert,kd_distilhubert,lightformer}, to accelerate learning and improve final performance \cite{KD_shao2023whisperkdq}.

\begin{figure}[t]
  \centering
  \includegraphics[width=0.75\linewidth, page=1,trim={11.41cm 7.4cm 10.94cm 5.05cm},clip]{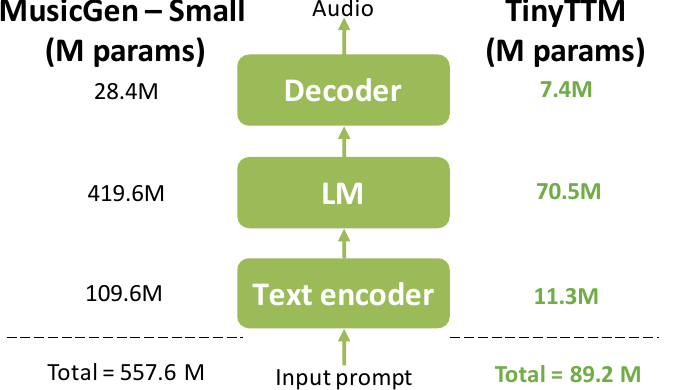}
  \vspace*{-5pt}\caption{MusicGen-Small vs. proposed TinyTTM.}
  \label{fig:main_diagram}\vspace{-22pt}
\end{figure}

Figure \ref{fig:main_diagram} summarizes this work, where we start from a pretrained MusicGen-Small model and through methods presented in this paper, reduce the model size from 557.6M parameters down to 89.2M parameters. This reduction process is achieved by individually optimizing each of the three components of the TTM model through a method that combines distillation and fine-tuning techniques using MusicBench \cite{mustango} dataset.
Contributions of this work are:
\begin{itemize}
    \item The creation of a tiny TTM system containing 89.2M parameters, for first time to the best of our knowledge, demonstrating performance comparable to larger systems and more suitable for deployment on resource-constrained devices.
    \item A knowledge distillation technique applying adaptive learning using stochastic loss sampling is proposed. In this method, we give new importance to different parts of the loss - like teacher loss, ground-truth loss, and model intermediate loss - with each step of distillation. This loss weight scheduling helps us perform better than existing methods where the weights given to each part of the loss is fixed. In contrast to fixed weighting, our method improves representation learning by adding uncertainty to how we pick the losses. 
    Two different distributions are explored to model the loss weight scheduling. 
\end{itemize}

\section{TinyTTM}
This section describes the MusicGen \cite{musicGen} components and the methods proposed to effectively reduce the model size.
\subsection{Encoder}
The encoder is used to create the embedding vector for conditioning the generator. 
This work uses T5 (Text-to-Text Transformer)  \cite{raffel2020exploring} following the structure proposed by MusicGen\cite{musicGen}. 

The T5 model, based on a Transformer encoder-decoder architecture, is offered in various sizes \cite{tay2022scale}. This paper focuses predominantly on T5-tiny (utilized in TinyTTM) and compares the performance with T5-base (implemented in MusicGen-small).
We evaluate the T5-tiny pretrained model\footnote{\url{https://huggingface.co/google/t5-efficient-tiny}} and fine-tune it on text containing music descriptions from MusicBench. In order to fine-tune the model we leverage the span-based MLM objective and cross-entropy loss \cite{raffel2020exploring}. During our investigation, we specifically explore the strategy of freezing the decoder layers during training, as these layers are not utilized during inference. Additionally, to further reduce the four layer T5-Tiny encoder  we explore dropping the last layers.

\subsection{Generative Transformer Language Model (LM)}

The Language Model (LM) in MusicGen \cite{musicGen} is trained using cross-entropy loss between the predicted output logits and the ground truth audio tokens generated by EnCodec. We define the student loss $L_{S}$ as:

\vspace*{-8pt}\begin{equation}
L_{S} = -\frac{1}{K} \sum_{k=1}^{K} \sum_{t=1}^{T} \sum_{c=1}^{C} y_{k,t,c} \log(p_{k,t,c})
\end{equation}

\noindent where $T$ are the temporal steps, $K$ the number of codebooks, $C$ the cardinality of each codebook, and $y_{t,c}$ and $p_{t,c}$ represent the ground truth and predicted probability of the $c^{th}$ token at time $t$, respectively. Batch dimension is ommited for simplicity.
 
 In this work, prior knowledge from a teacher model is leveraged to guide the student model. This is accomplished through the utilization of both the teacher LM outputs, as well as the hidden representations within the transformer architecture. We define the teacher loss $L_{T}$ as KL divergence based \cite{kd_literature_review_main,KD_shao2023whisperkdq}:

\vspace*{-8pt}\begin{equation}
 L_{T} = -\frac{1}{K} \sum_{k=1}^{K} \sum_{t=1}^{T} \sum_{c=1}^{C} q_{k,t,c} \log\left(\frac{q_{k,t,c}}{p_{k,t,c}}\right)
\end{equation}

\noindent where $q_{k,t,c}$, $p_{k,t,c}$ are the predicted probability of the $c^{th}$ token at time $t$ for the $k^{th}$ codebook according to teacher and student model, respectively. The  teacher loss $L_{T}$ is the averaged KL divergence across all $K$ codebooks.
Additionally, the teacher intermediate transformer layer outputs $\hat{y}_{T}$ are being utilized as guidance to align the student corresponding outputs  $\hat{y}_{S}$:

\vspace*{-15pt}\begin{equation}
L_{MSE}(\hat{y}_{T}, \hat{y}_{S}) = \frac{1}{K_{tr}} \sum_{k=1}^{K_{tr}} |\hat{y}_{T,m(k)} - \hat{y}_{S,k}|^2
\end{equation}

\noindent where $K_{tr}$ denotes the number of student transformer layers, and $|\cdot|^2$ signifies the squared Euclidean norm. The additional subscript in $\hat{y}_{T,m(k)}$ and $\hat{y}_{S,k}$ represent the layer index and $m(k)$ represents a function that maps the teacher transformer layer ids to the corresponding ones from the student, based on the layer matching strategy that is employed.
The final loss is given by the weighted sum of the aforementioned loss terms:

\vspace*{-8pt}\begin{equation}
    L_{LM} = l_{s} \cdot L_{S} + l_{t} \cdot L_{T} + l_{m} \cdot L_{MSE}
    \label{eq:LLM}
\end{equation}

\noindent where $l_s$, $l_t$ and $l_m$ are different factors applied to each of the loss term to align the magnitudes to the same range.

\subsubsection{Weight Transferring}

Weight transferring technique employs the pre-trained weights of a larger model to enhance the initialization of student models \cite{kd_literature_review_main, lightformer}. 
Random initializations select different modes within the function space, compared to initializations with subspace sampling, which is trying to cluster within a singular mode \cite{fort2019deep}. The subspace initilizations are far from the initial weight space but exhibit similarities in the function space, resulting in a less diverse optimal space \cite{fort2019deep}.

In this work, two weight initialization methods are compared: random initialization and weight transferring. Since the student model is characterized by a significantly reduced number of layers compared to the teacher model, we adopt an equidistant selection strategy \cite{kd_literature_review_main, kd_distilbert, lightformer} of teacher layers to transfer.

\vspace{-2mm}
\subsubsection{Dynamic Loss Weight Scheduling}
\label{lm:dynamic_loss_weight}

The adaptation of function space from the teacher model to the compressed student model requires constant leveraging of the weights among the prior knowledge representation and learned knowledge representation at a given time. This is why weighting different aspects in the loss function contributes crucially to the model optimization process \cite{Shimodaira2000ImprovingPI, jiang2015self}. While random weighting methods are theoretically stochastic versions of equal weighting approaches, they (random weighting) exhibit a higher likelihood of avoiding local minima \cite{NEURIPS2019_e58cc5ca, fang2020rethinking}. Consequently, this characteristic leads to improved generalization performance compared to equal weighting methods \cite{NEURIPS2019_e58cc5ca}. Although, the weights in the loss functions are crucial in the initial stage of the training, they lose importance in the later stages of training \cite{pmlr-v97-byrd19a}. Therefore, in this work the goal is to create randomness in the gradient noise while training within the same mini-batch, which in turn implicitly forces the model to learn a generalised representation. 
We explore modifying the contribution of each loss term in (\ref{eq:LLM}) dynamically during each training step, as follows:

\vspace*{-12pt}\begin{equation}
    L_{LM} = a_{1} \cdot l_{s} \cdot L_{S} + a_{2} \cdot l_{t} \cdot L_{T} + a_{3} \cdot l_{m} \cdot L_{MSE}
    \label{eq:LLM_weighted}
\end{equation}

\noindent where $a_{i}, i \in \{1,2,3\}$ s.t. $\sum_i a_i = 1$ are drawn from a distribution created with Algorithm \ref{algo:method2} or Algorithm \ref{algo:smith}, which is introduced in \cite{smith2004sampling}. %

\vspace*{-10pt}
\noindent %
\begin{minipage}{.22\textwidth} %
\vspace{-33.7pt}

\begin{algorithm}[H]
\scriptsize{
\caption{Sampling \\ Strategy 1 - S1}\label{algo:method2}
\begin{algorithmic}[1] %
\For{$i\in\{1,2,3\}$}
\State$a_{i}\sim uniform(0,1)$
\EndFor
\State $S \gets \sum_{j=1}^{3}{a_{j}}$
\For{$i \in \{1,2,3\}$}
\State $a_{i} \gets \frac{a_{i}}{S}$ 
\EndFor 
\end{algorithmic}
}
\end{algorithm}
\end{minipage}%
\hfill
\begin{minipage}{.24\textwidth}
\begin{algorithm}[H]
\scriptsize{
\caption{Sampling \\Strategy 2 - S2\cite{smith2004sampling}}\label{algo:smith}
\begin{algorithmic}[1] %
\State$n \gets 3$
\State $M \gets 2^{32}$
\State $a \gets $ Generate $n-1$ distinct
values $a_{i}$ from $[1, M-1]$
\State $a \gets$ sort($a$)
\State $a \gets$ insert($a, 0, 0$)
\State $a \gets$ append($a, M$)
\State $a \gets$ diff($a$)
\For{$i \in \{1, \ldots, n\}$}
\State $a_i \gets \frac{a_i}{M}$ 
\EndFor 
\end{algorithmic}
}
\end{algorithm}
\end{minipage}

 \vspace*{8pt}Training with randomly initialized loss weights $a_i$ introduces gradient noise within the mini-batch, which enables further generalised learning for the model.

\subsection{Decoder}
\label{decoder_section}
The decoder of the EnCodec model \cite{encodec} is used which is a neural audio codec that compresses the audio into a set of tokens, resulting in a lower bitrate and it features a streaming convolutional-based encoder-decoder architecture, with sequential modeling applied to the latent representation. The encoder includes four 1D convolution blocks with 64 channels, including residual units and down-sampling layers, as well as two-layer LSTM and a final 1D convolution. The decoder, mirroring the encoder design in reverse, transforms the latent representation into the output audio. Each downsampling step doubles the number of channels to enhance the model capabilities.

The Encodec distillation approach proposed is a combination of losses using the ground truth labels and the teacher output. Concretely, let $x$ be the ground truth audio, $\hat{x}_S$ be the output generated by the student model, $\hat{x}_{T}$ be the generated audio by the teacher model. The reconstruction distillation loss in the time domain is given by:
\vspace{-1mm}
\vspace{-4pt}\begin{equation}
    l_t(\hat{x}_S,\hat{x}_{T}) = ||\hat{x}_S-\hat{x}_{T}||_1
\end{equation}
In the reconstruction distillation loss in the frequency domain is the linear combination between L1 and L2 losses over the Mel-spectrogram:
\vspace{-8pt}\begin{multline}
  l_f(\hat{x}_S, \hat{x}_{T}) = \frac{1}{| \alpha | \cdot | s |} \sum_{\alpha_i \in \alpha} \sum_{i \in e} || S_i(\hat{x}_S) - S_i(\hat{x}_{T}) ||_1\\
  + \alpha_i || S_i(\hat{x}_S) - S_i(\hat{x}_{T}) ||_2  
\end{multline}

\noindent The adversarial distillation loss for the generator is:
\vspace{-4pt}\begin{equation}
    l_g(\hat{x}_S,\hat{x}_{T}) = \frac{1}{K} \sum_{k} \max(0, 1 - D_k(\hat{x}_{T}))
\end{equation}
where $K$ is the number of the discriminators utilized.
The adversarial distillation feature matching loss for the generator is formulated as:
\vspace{-3mm}
\vspace{-4pt}\begin{equation}
    l_{feat}(\hat{x}_S, \hat{x}_{T}) = \frac{1}{KL} \sum_{k=1}^{K} \sum_{l=1}^{L} \frac{|| D_{k}^L(\hat{x}_{S}) - D_{k}^L(\hat{x}_{T}) ||_1} {\text{mean} \left( || D_{k}^L(\hat{x}_{S}) ||_1 \right)} 
\end{equation}

\noindent The final loss $L_G$ used to distill the teacher EnCodec is:

\vspace{-3mm}
\vspace{-4pt}\begin{align}
    L_{G} =  &  \lambda_t\cdot(l_t(x,\hat{x}_S) + l_t(\hat{x}_S,\hat{x}_{T})) +  \lambda_f\cdot(l_f(x, \hat{x}_S) \\
    & + l_f(\hat{x}_{S}, \hat{x}_{T})) + \lambda_g\cdot(l_g(x,\hat{x}_S) + l_g(\hat{x}_S,\hat{x}_{T})) \nonumber \\
    & + \lambda_{feat}\cdot(l_{feat}(x, \hat{x}_{S})+l_{feat}(\hat{x}_S, \hat{x}_T)) + \lambda_w \cdot (l_w(w)) \nonumber
\end{align}

\noindent where $\lambda_t, \lambda_f, \lambda_g, \lambda_{feat}, \lambda_w $ are the scalars to balance between the terms and their values, which are 0.1, 2, 4, 4 and 0.1. The $l_w(w)$ loss measures the Euclidean distance between the current residual and its nearest codebook entry added over all residual steps in the batch. It should be emphasized that the discriminator was trained using a combination of teacher and student outputs, thus the discriminator loss was updated as:
\vspace{-1mm}
\vspace{-8pt}\begin{align}
   & L_d(x,\hat{x}_S,\hat{x}_{T})  = \frac{1}{K}\sum_{k=1}^K 2\cdot (\max(0,1-D_k(x))  \\
    & \quad \quad \quad \quad  + \max(0,1+D_k(\hat{x}_S))) + \max( 0,1+D_k(\hat{x}_{T})) \nonumber 
\end{align}
\noindent Distillation in the EnCodec is only implemented in the decoder to maintain the same encoder and quantizer as the teacher, preventing potential token mismatch in the MusicGen token generation model. The number of filters in the convolution kernels is decreased to reduce the number of parameters. Although experiments involving a single LSTM layer, without altering the filter numbers were conducted, it led to a relatively high size of the decoder ($\approx$20M parameters).

\vspace{-3mm}

\section{Evaluation setup}
The experiments in this work are performed on the MusicBench database \cite{mustango}. MusicBench is derived from MusicCaps \cite{musicLM} by augmenting the audio with musically meaningful techniques (semitone pitch shifts, tempo changes and volume changes) and enhancing the captions. The total number of tracks for training is 52,768 leaving 400 samples for test.

We use two objective evaluation metrics to measure the performance of the music compute with the AudioLDM eval toolkit\footnote{\url{https://github.com/haoheliu/audioldm_eval}}: Fréchet Audio Distance (FAD) \cite{fad} and Kullback–Leibler (KL) divergence. The FAD indicates the similarity of the generated audio set with the target audio set in terms of the VGGish \cite{vggish} features distribution. The KL is instead measured between the generated sample and target sample features extracted with PANNs model\cite{panns} and then  averaged for the entire set.

\section{Experimental Analyses}
In this section we analyse the performance of each of the three TinyTTM modules individually, as well as the performance of the TinyTTM on the MusicBench dataset.

\subsection{Encoder analysis}\label{sec:encoderAnalysis}
In Table \ref{tab:T5_results} we present the music generation performance of different encoder configurations. The evaluation is carried out by replacing the text encoder and fine-tuning the LM of the MusicGen pipeline. The LM configuration used for these experiments is the Variant 1 in Section  \ref{sec:lm_analysis}, trained without ground truth labels, and the default decoder. The first two models, in Table \ref{tab:T5_results}, are the default ones without any modification. The last four are fine-tuned on the textual descriptions of MusicBench dataset exploiting span-based MLM objective and cross-entropy loss  \cite{raffel2020exploring}. In the last two experiments we remove the last one and last two encoder layers (out of four layers) respectively with the aim of decreasing the size.

The scores in Table \ref{tab:T5_results} represent the average of five distinct generations generated using different seeds. This table highlights the considerable performance gap between the default T5-base and T5-tiny in TTM. Fine-tuning the T5-tiny model leads to enhanced results, reducing both the FAD and KL scores. However, the T5-base default configuration still achieves lower scores at the cost of larger model size (approximately tenfold). Fine-tuning the T5-tiny with a frozen decoder does not yield improvements, potentially due to the limited trainable parameters. Finally, T5-tiny models fine-tuned with two or three encoder layers results in a slight deterioration compared to the initial T5-tiny model with four layers, suggesting a trade-off between model parameters and effectiveness.

\vspace*{-8pt}\begin{table}[h!]
\scriptsize
  \caption{Performance on the different T5 configurations. FT: fine-tuned on MusicBench dataset (train set).}\vspace*{-10pt}
  \label{tab:T5_results}
  \centering
  \begin{tabular}{cccc}
    \toprule
    \textbf{Method} & \textbf{FAD} & \textbf{KL} & \textbf{\#params} \\
    \midrule
     T5-base default & 3.83 & 1.34 & 109.6M \\
     T5-tiny default & 4.14 & 1.64  & 11.3M \\
     \hline
     T5-tiny FT & {\bf 3.92} & {\bf 1.41} & 11.3M \\
     T5-tiny FT frozen decoder & 4.17 & 1.50 & 11.3M \\
     T5-tiny FT 3 encoder layers & 4.02 & 1.48 & 10.6M \\
     T5-tiny FT 2 encoder layers & 4.00 & 1.46 & 9.8M \\
    \bottomrule
  \end{tabular}
\end{table}\vspace*{-15pt}

\subsection{Transformer Language Model analysis} \label{sec:lm_analysis}

\noindent Two variants of compressed transformers have been employed in the following experiments. \textbf{Variant 1} (V1) includes a reduced number of transformer layers, but preserves the dimensionality of the teacher, to perform weight transfer. \textbf{Variant 2} (V2) comprises more transformer blocks cascade, but employs smaller  dimensionality, in order to maintain the number of parameters with Variant 1. A more detailed presentation of each variant configuration is shown in Table \ref{tab:transformer_variants}.

\begin{table}[ht]
\scriptsize
\centering
\caption{LM Variants details. Teacher is MusicGen-Small.}\vspace*{-10pt}
\label{tab:transformer_variants}
\begin{tabular}{lccc}
\hline
\textbf{Feature} & \textbf{Variant 1} & \textbf{Variant 2} & \textbf{Teacher}\\
\hline
Layers & 4 & 7 & 24 \\
Heads & 16 & 8 & 16 \\
Transformer dim. & 1024 & 720 & 1024 \\
Parameters & 84.71M & 70.5 M & 419.6M \\
\hline\vspace*{-20pt}
\end{tabular}
\end{table}

\subsubsection{Teacher LM Model Fine-Tuning}
The MusicGen-small teacher model  has been trained exclusively with audio samples that contain music without the presence of any vocals\footnote{\url{https://github.com/facebookresearch/audiocraft/blob/main/model_cards/MUSICGEN_MODEL_CARD.md}}. 
Since MusicBench contains vocals, we also fine-tune the pretrained model on the MusicBench training set before distillation to enable the model to generate vocals. 
In Table \ref{tab:teachers_performance_overall} we analyse the performance on MusicBench entire test set (400 samples) and two subsets from it: \textbf{V} (192 samples) containing reference vocals in the text description and \textbf{NV} (208 samples) with no reference to vocals. This classification has been performed with the help of Mixtral 8x7B\footnote{\url{https://huggingface.co/mistralai/Mixtral-8x7B-v0.1}} LLM \cite{mixtral}. %

\vspace*{-8pt}
\begin{table}[ht]
\scriptsize
\centering
\caption{ Comparative results regarding FAD, KL metrics on the Musicbench Test A set for the two candidate teacher models.}\vspace*{-10pt}
\label{tab:teachers_performance_overall}
\begin{tabular}{lcccccc}
\hline
& \multicolumn{2}{c}{\textbf{Entire set}} & \multicolumn{2}{c}{\textbf{V subset}} & \multicolumn{2}{c}{\textbf{NV subset}} \\
\textbf{Model} & \textbf{FAD} & \textbf{KL} & \textbf{FAD} & \textbf{KL} & \textbf{FAD} & \textbf{KL}\\
\hline
Teacher \cite{musicGen}  & 3.85 & 1.33& 5.84 & 1.46  & 3.93 & 1.29 \\
Teacher FT  & 2.73 & 1.20& 3.19 & 1.24 & 3.37 & 1.23 \\
\hline
\end{tabular}
\end{table}

Table \ref{tab:teachers_performance_overall} shows that the FAD and KL of the fine-tuned model (Teacher FT) have improved on the \textbf{V} subset, as well as on the overall test set. While the differences are minor on the \textbf{NV} set, where the fine-tuned model is still comparable to the pretrained one, in terms of FAD and KL.

\subsubsection{Student LM Knowledge Distillation}
In Table \ref{tab:lm_ablations} we present an ablation study on different distillation configurations and show the impact on FAD and KL.
The fine-tuned teacher model on MusicBench demonstrates a notable improvement over the pre-trained MusicGen-Small model. 
Compressed models V1 and V2  further improve the performance under different training regimes.
Overall the V2 architecture shows lower FAD and KL. The incorporation of teacher loss and mse loss results in the most favorable outcomes in terms of FAD and KL, while the application of loss sampling strategies further optimizes the performance.

\subsection{Decoder analysis} \label{sec:decoder_analysis}
Table \ref{tab:EnCodec_results} illustrates the FAD and KL achieved with EnCodec. The target in the evaluation is the input audio to the EnCodec encoder which is compared to the output generated by the EnCodec decoder.
The weight factor is applied to $\lambda_t, \lambda_f, \lambda_g, \lambda_{feat}, \lambda_w $ which are the scalar coefficients used for balancing the gradients. Based on the results, the distilled EnCodec is able to achieve better performance compared to the fine-tuned (Tiny EnCodec FT) for the FAD metric, while there is a slight drop (0.01) in the KL metric. Adding a weight factor to the scalar coefficients, when distilling the model, has significantly improved the model's performance in terms of FAD, while both models maintain the same performance for the KL metric. The experiment with the weight factor equal to 0.75 provided the best results among the other experiments. 
As a consequence, as decoder of the TinyTTM, the distilled EnCodec with weight factor equal to 0.75 was selected.

\begin{table}[ht!]
\scriptsize
\centering
\caption{Performance (mean $\pm$ standard deviation) of distilled LM with various setups. \textbf{H}: Student loss, \textbf{S}: Teacher loss, \textbf{mse}: Intermediate MSE loss. $\boldsymbol{S_{i}}$: Sampling strategy $i$ from \ref{lm:dynamic_loss_weight}.}\vspace*{-10pt}
\label{tab:lm_ablations}
\begin{tabular}{lccc}
\hline
\textbf{Variant} & \textbf{Init. Strat.} & \textbf{FAD} & \textbf{KL} \\
\hline
Teacher \cite{musicGen} & Random & 3.85 {\footnotesize $\pm$ 0.10} & 1.38 {\footnotesize $\pm$ 0.01} \\
Teacher FT & Weight Copy & 2.73 {\footnotesize $\pm$ 0.06} & 1.28 {\footnotesize $\pm$ 0.01} \\
\hline
V1 H & Random & 4.08 {\footnotesize $\pm$ 0.11} & 1.61 {\footnotesize $\pm$ 0.04}  \\
V1 S & Random & 3.62 {\footnotesize $\pm$ 0.05} & 1.52 {\footnotesize $\pm$ 0.02} \\
V1 H & Weight copy & 4.17 {\footnotesize $\pm$ 0.07} & 1.64 {\footnotesize $\pm$ 0.01}  \\
V1 S & Weight copy & 4.01 {\footnotesize $\pm$ 0.06} & 1.51 {\footnotesize $\pm$ 0.02} \\
V1 H/mse & Random & 3.80 {\footnotesize $\pm$ 0.11} & 1.54 {\footnotesize $\pm$ 0.02} \\
V1 S/mse & Random & 3.85 {\footnotesize $\pm$ 0.15} & 1.54 {\footnotesize $\pm$ 0.03} \\
V1 H/S/mse & Random & 3.61 {\footnotesize $\pm$ 0.01} & 1.55 {\footnotesize $\pm$ 0.03} \\
\hline
V2 H & Random & 4.20 {\footnotesize $\pm$ 0.05} & 1.62 {\footnotesize $\pm$ 0.01} \\
V2 S & Random & 3.76 {\footnotesize $\pm$ 0.09} & 1.55 {\footnotesize $\pm$ 0.01} \\
V2 H/mse & Random & 3.83 {\footnotesize $\pm$ 0.01} & 1.57 {\footnotesize $\pm$ 0.01} \\
V2 S/mse & Random & 3.47 {\footnotesize $\pm$ 0.12} & 1.48 {\footnotesize $\pm$ 0.04} \\
V2 H/S/mse & Random & 3.45 {\footnotesize $\pm$ 0.03} & 1.51 {\footnotesize $\pm$ 0.03} \\
\hline
V2 H/S/mse/$S_{1}$ & Random & {\bf 3.40} {\footnotesize $\pm$ 0.02} & 1.49 {\footnotesize $\pm$ 0.01} \\
V2 H/S/mse/$S_{2}$ & Random & 3.53 {\footnotesize $\pm$ 0.03} & {\bf 1.43} {\footnotesize $\pm$ 0.02} \\
\hline
\end{tabular}
\vspace*{-6pt}
\end{table}
\vspace*{-10pt}\begin{table}[ht!]
  \caption{Performance on EnCodec model. w.f.: weight factor}\vspace*{-10pt}
  \label{tab:EnCodec_results}
  \scriptsize
  \centering
  \begin{tabular}{cccc}
    \toprule
    \textbf{Method} & \textbf{FAD} & \textbf{KL} & \textbf{\#params} \\
    \midrule
    Teacher EnCodec & 3.84 & 1.15 & 28.4M \\\hline
    Tiny EnCodec FT & 3.21 & 0.88 & 7.43M \\
    Distilled EnCodec & 3.17 & 0.89 & 7.43M \\
    Distilled EnCodec (w.f. = 0.75) & {\bf 2.87} & {\bf 0.88} & 7.43M \\
    Distilled EnCodec (w.f. = 1.25) & 3.15 & 0.89 & 7.43M \\
    \bottomrule\vspace*{-20pt}
  \end{tabular}
\end{table}

\subsection{\vspace*{-4pt}TinyTTM performance}

We combined the best performing TTM components in the previous sections to create TinyTTM. 
Table \ref{tab:lm_complete_results_2} illustrates a comparison between the two candidate teacher models and the best performing TinyTTM models, in terms of FAD, KL and total parameter count. It is observed that the compressed models are better than the teacher model in terms of FAD and KL, however do not achieve the performance of the fine-tuned teacher.
The size compression is up to $\times 6.25$ with $\times2.8$ latency reduction.\footnote{Comparison using a single Nvidia A10 GPU.}
Although TinyTTM achieves a low FAD and KL, some of the generated tracks have noticeable issues: beat synchronization, singing voice not created or not sounding like voice.
This can be due to the low audio diversity in MusicBench as this dataset is created from only 5k tracks.
\vspace{-7.2pt}
\begin{table}[h!]
  \caption{Performance comparison (for 3 runs with different seeds) between the proposed \textbf{TinyTTM} model and MusicGen-Small  on the MusicBench test set A.}\vspace*{-10pt}
  \label{tab:lm_complete_results_2}
  \scriptsize
  \centering
  \begin{tabular}{cccc}
    \toprule
    \textbf{Model/Method} & \textbf{FAD} & \textbf{KL} & \textbf{\#params}  \\
    \midrule
    Teacher \cite{musicGen}& 3.85 $\pm 0.11$  & 1.33 $\pm 0.01$ & 557.6M  \\
    Teacher FT  & 2.73 $\pm 0.05$ & 1.20 $\pm 0.01$ & 557.6M \\
    \midrule
    V2 H/S/mse/$S_{1}$ & 3.85 $\pm 0.06$ & 1.39 $\pm 0.02$ & 89.2M \\
    V2 H/S/mse/$S_{2}$ & {\bf 3.66 $\pm 0.14$ } & 1.32 $\pm 0.01$ & 89.2M  \\
    \bottomrule
  \end{tabular}\vspace*{-2pt}
\end{table}

\vspace{-6mm}
\section{Conclusions}
We have presented \textbf{TinyTTM}, a TTM model that operates within a parameter-constrained framework. We highlighted the limitations of employing only ground truth labels for training as outlined in \cite{musicGen}, and discussed the enhancements achieved through KD techniques applied to each of the TTM components. With a 89.2M parameters we are able to achieve FAD of 3.66 and KL of 1.32, which is lower than the model in \cite{musicGen}, but not as low as the fine-tuned counterpart.
Future directions include expanding the exploration of KD techniques across broader and more diverse datasets to solve observed issues when training with MusicBench (like beat synchronization or voice distortion
) and thus reducing the performance gap between TinyTTM and MusicGen-Small fine-tuned.

\clearpage
\bibliographystyle{IEEEtran}
\bibliography{mybib}

\end{document}